\documentclass[10pt]{article}
\usepackage{graphics, epsfig,colordvi,afterpage}
\usepackage{times}
\usepackage{multirow}
\usepackage{palatino}
\usepackage{amsmath,amssymb}
\usepackage{caption}
\usepackage{placeins}
\usepackage{etoolbox}
\def\AnnotatedKeys{Zhaoping_iScience2025}

\makeatletter
\pretocmd{\@bibitem}{\def\currentbibkey{#1}}{}{}
\pretocmd{\@lbibitem}{\def\currentbibkey{#2}}{}{}
\makeatother

\newcommand{\printannotation}{%
  \ifinlist{\currentbibkey}{\AnnotatedKeys}{%
    \par\smallskip\small\itshape
    \csname annotation@\currentbibkey\endcsname
  }{}%
}

\makeatletter
\apptocmd{\end@bibitem}{\printannotation}{}{}
\makeatother

\usepackage{wrapfig}
\usepackage{xcolor}

\mathchardef\mhyphen="2D
\input{epsf}

\def\be{\begin{equation}}
\def\ee{\end{equation}}
\def\bea{\begin{eqnarray}}
\def\eea{\end{eqnarray}}

\def\br{{\bf r}}

\def\bibsinglebull{$\phantom{\bullet}\bullet$~}
\def\bibdoublebull{$\bullet\bullet$~}

\makeatletter
\renewenvironment{thebibliography}[1]
     {\section*{\refname}%
      \@mkboth{\MakeUppercase\refname}{\MakeUppercase\refname}%
      \list{%
      \phantom{$\bullet$~}%
      \ifnum\c@enumiv=47\bibdoublebull\else     
      \ifnum\c@enumiv=14\bibdoublebull\else     
      \ifnum\c@enumiv=12\bibsinglebull\else     
      \ifnum\c@enumiv=42\bibdoublebull\else   
      \ifnum\c@enumiv=51\bibsinglebull\else   
      \ifnum\c@enumiv=54\bibdoublebull\else   
      \ifnum\c@enumiv=63\bibsinglebull\else   
      \ifnum\c@enumiv=72\bibsinglebull\else   
    \phantom{$\bullet\bullet$~}\fi\fi\fi\fi\fi\fi\fi\fi
      \@biblabel{\@arabic\c@enumiv}}%
           {\settowidth\labelwidth{\@biblabel{#1}}%
            \leftmargin\labelwidth
            \advance\leftmargin\labelsep
            \@openbib@code
            \usecounter{enumiv}%
            \let\p@enumiv\@empty
            \renewcommand\theenumiv{\@arabic\c@enumiv}}%
      \sloppy
      \clubpenalty4000
      \@clubpenalty \clubpenalty
      \widowpenalty4000%
      \sfcode`\.\@m}
     {\def\@noitemerr
       {\@latex@warning{Empty `thebibliography' environment}}%
      \endlist}
\makeatother

\begin{document}
\baselineskip = 16 pt
\footskip = 0.75 in
\setlength{\parindent}{ 0.33 in}

\makeatletter

{\Large \bf What are the functions of primary visual cortex (V1)?}

\centerline{Li Zhaoping}

\centerline{University of T\"ubingen, Max Planck Institute for Biological Cybernetics, T\"ubingen, Germany}
\centerline{email: li.zhaoping@tuebingen.mpg.de}
\centerline{in Press for {\it Current Opinions in Neurobiology} 2026}

\subsection*{Abstract:} 

Although Hubel and Wiesel established decades ago how individual V1 neurons
transform retinal inputs, functions of V1 as a whole are being discovered only recently.
First, V1 acts as a motor cortex for exogenously guiding saccades by constructing a bottom-up saliency map of the visual field.
Second,  V1 initiates a processing bottleneck: a massive reduction of visual information
begins at its output to downstream areas.
Third, downstream recognition is limited by impoverished information,
V1 supports ongoing recognition by providing additional information queried by
top-down feedback from downstream areas, directed predominantly to central visual field representations.
These V1 functions underpin a framework in which vision is mainly looking and seeing through the bottleneck.
Looking selects a fraction of visual information into the bottleneck, largely by saccades that center selected
contents at gaze. Seeing recognizes the selected contents.
Looking and seeing rely mainly on processing in the peripheral and central visual fields.

\section*{Highlights} 

\begin{itemize}
\item Primate V1 serves vision's main function of looking and seeing through a bottleneck.
\item It creates a bottom-up saliency map to exogenously guide looking via gaze shifts.
\item It initiates an information bottleneck at V1's output to downstream brain areas
\item It supports feedback queries mainly directed to the central visual field
\item The queried information aids ongoing visual recognition in light of the bottleneck
\end{itemize}

\section*{Keywords}

primary visual cortex, gaze shifts, saliency map, bottleneck, top-down feedback, selection, recognition,
the looking-and-seeing framework.  

%

\section*{Introduction}

\begin{figure}[h!!!]
\begin{center}
\includegraphics[width=150mm]{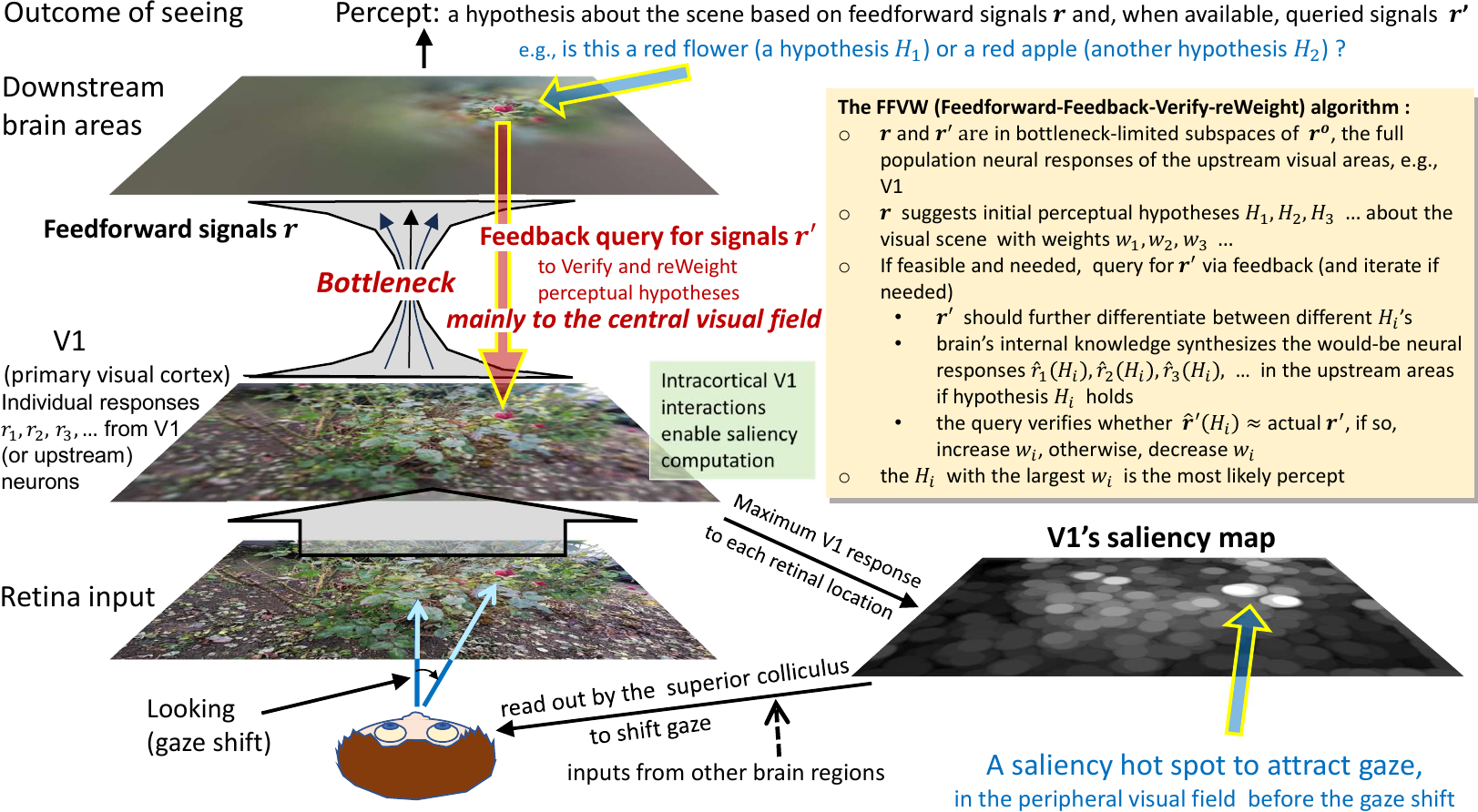}
\end{center}
\caption{\label{fig:FigRose}
Functional roles of the primary visual cortex V1. 
V1 first creates a bottom-up saliency map that
exogenously guides gaze shifts (looking),  selecting a 
fraction of visual input for the brain's processing bottleneck and centering
it at the point of gaze.
V1 also initiates this bottleneck, limiting 
downstream information and thereby making recognition (seeing) nontrivial.
To support ongoing recognition, V1 supplies
additional information when queried by feedback from downstream areas along the visual pathway. 
This feedback predominantly targets the central visual field, 
which is specialized for seeing.  In contrast, the peripheral visual field is specialized for looking, 
determining where to shift gaze.
}
\end{figure}

V1 is not alone in that we remain ignorant about 
its function
decades after we have known its neural properties\cite{HubelWiesel1962, 
CarandiniEtAl2005, OlshausenField2006}. 
Before patient HM, the hippocampus's critical role in memory remained  unknown long after its anatomy 
was characterized.  V1 is clearly essential for vision, but so is the retina.  
What, then,  is V1's special function?

The emerging answer is tied to the question of what vision is.
Our brain faces an information processing bottleneck,
constrained by metabolic energy \cite{AttwellLaughlin2001}, space for neurons and wiring, 
and time for (e.g.,) complex sensory inference \cite{ZhaopingBook2014}.
Often termed the attentional bottleneck, it allows only a tiny fraction of visual inputs, 
largely near the center of gaze, to reach recognition.
Vision is therefore not only seeing, but also looking \cite{ZhaopingBook2014}:
looking selects the fraction allowed to enter the bottleneck (cf. attentional selection), 
typically by gaze shifts that center the selected inputs on the fovea;
and seeing (recognizing) decodes and discriminates this fraction.    
This looking-and-seeing framework has developed alongside investigations of V1's function,
beginning with the V1 Saliency Hypothesis (V1SH) in the late 1990s 
\cite{LiPNAS1999, LiTICS2002}, and progressing to the central-peripheral dichotomy (CPD) 
theory from the 2010s \cite{ZhaopingFFVW2017, ZhaopingNewFramework2019}. 
This framework is supported by empirical findings, particularly the most recent ones
and various experimental confirmations of falsifiable predictions.

First, V1 serves looking by guiding gaze shifts exogenously.
According to V1SH, it creates a map of salience, defined as the strength of a visual location to 
attract attention by external visual inputs (Figure \ref{fig:FigRose}).  
This map is read out by the superior colliculus (SC), which specifies saccades 
to salient locations.  Second, to reduce cost, the bottleneck begins immediately at V1's output to downstream 
stages. Third, to aid ongoing recognition (seeing) from the limited information 
admitted into the bottleneck, feedback from downstream stages 
queries upstream stages such as V1 for additional relevant information; 
according to CPD,  this feedback is largely 
restricted to the central visual field to reduce cost \cite{ZhaopingNewFramework2019}.
Accordingly, saccades to any scene location enable a query 
to that location.  Apparently, evolutionary optimization had led to the CPD 
in this looking-and-seeing framework  to balance computation and costs.

\begin{figure}[ttttt!!!]
\begin{center}
\includegraphics[width=150mm]{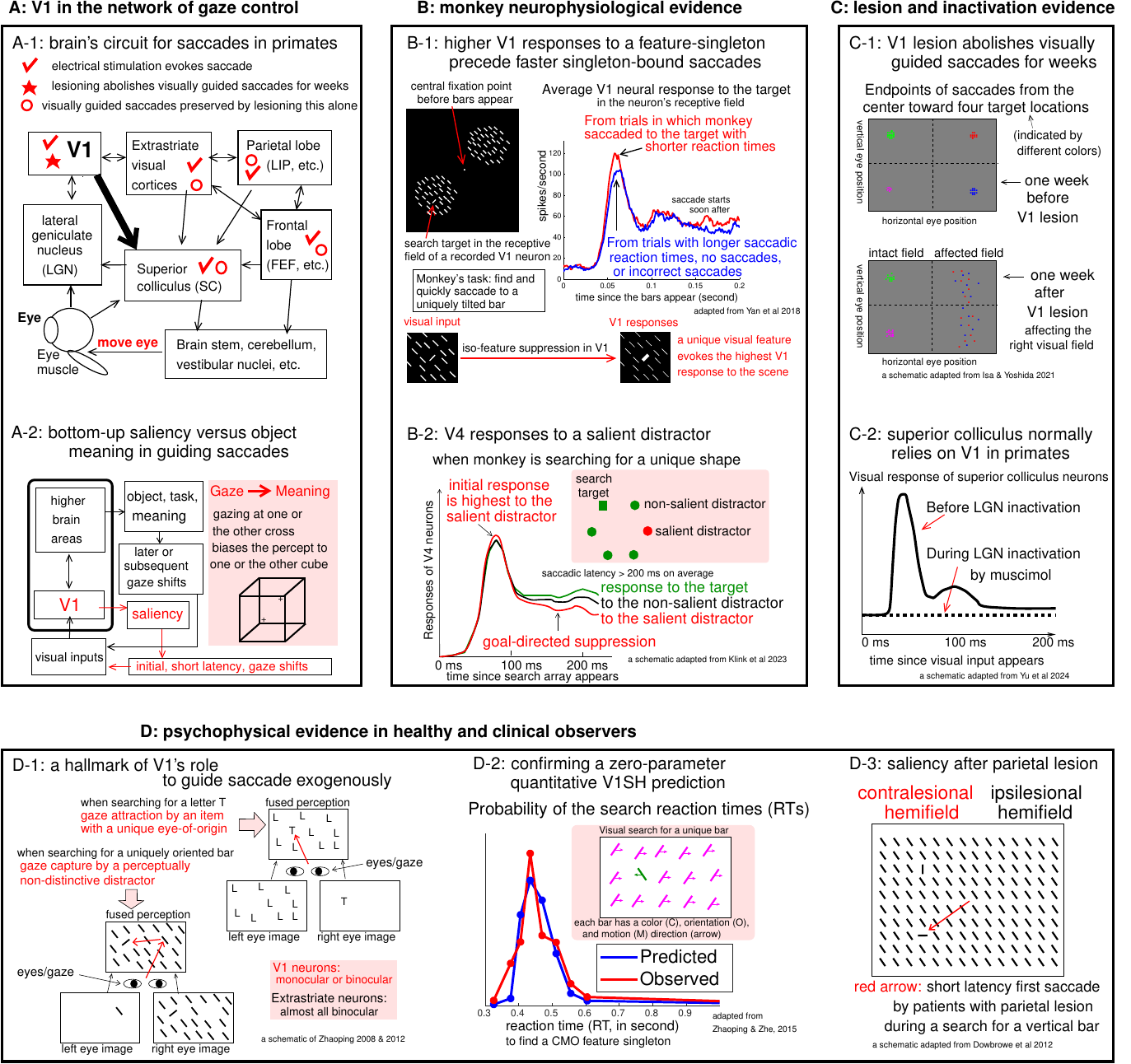}
\end{center}
\caption{\label{fig:FigV1Saliency}
\small V1 exogenously guides saccades. 
A: V1 in a network of brain areas for guiding gaze and attention.
A-1: in primates, electric stimulation of retinotopic locations in V1, 
extrastriate cortices, parietal cortex, superior colliculus (SC), or frontal eye field (FEF)
evokes saccades, but only V1 lesions abolish visually guided saccade for weeks.
A-2: a view on how bottom-up saliency (computed by V1) and object meaning jointly guide saccades.
B: neurophysiological evidence.
B-1: during a monkey's search for an orientation singleton bar, faster saccades 
to the target are typically preceded by higher initial responses of V1 neurons 
to the target.  
Iso-feature suppression makes nearby V1 cells tuned to similar features suppress
each other.  In this example, iso-orientation suppression makes the V1 neural 
response to the target typically the strongest response to this image.
B-2: in a visual search for a unique shape (randomly a square among disks or vice versa), 
V4 neural response to a salient distractor (randomly, red among green or vice versa) 
is initially significantly higher (somewhat exaggerated schematically for better 
visualization) than the responses to the target and other non-targets, 
before it is suppressed \cite{KlinkEtAl2023}.
C: lesion evidence.
C-1: after monkeys had learned to saccade to a light spot in any of the four corners
of a display for rewards, V1 lesions abolished visually guided --- but spared memory-guided --- saccades  
in the affected hemifield\cite{IsaYoshida2021}.
C-2: in monkeys, muscimol inactivation of LGN abolishes visual responses 
in SC neurons\cite{YuEtAl2024}. 
D: behavioral evidence.
D-1: V1 is the only cortical area to represent eye-of-origin accurately.
Uniqueness in eye-of-origin is perceptually invisible but strongly 
attracts gaze \cite{Zhaoping2008OcularSingleton, ZhaopingPeripheral2024}.
D-2: with no free parameters, V1SH accurately predicts a distribution of reaction times (RTs) 
to find a bar unique in color (C), orientation (O), and motion direction (M) 
from RTs to singleton bars unique in one or two of these three feature dimensions 
\cite{ZhaopingZhe2015}.
D-3: patients with parietal lesions retain saliency effects in the affected hemifield
in their short latency first saccades, which are more likely directed to a salient 
distractor in a visual search \cite{DombroweEtAl2012}.
}
\end{figure}
\FloatBarrier

\section*{V1 as a motor cortex to guide saccades exogenously for visual input selection}

By CPD, V1's role in looking is by its representation of the peripheral visual field, which we focus on in this section.
Like the conventional motor cortices, 
which project to muscles indirectly via spinal cord and brain stem, 
V1 projects indirectly to the eye movement muscles 
through SC, which is part of the brain stem. 
In primates,  electrical stimulation of V1, 
SC, frontal eye fields (FEF), extrastriate cortical areas (e.g.,V2 and V4), 
and parietal areas (e.g., LIP) can elicit eye movements 
(Fig. \ref{fig:FigV1Saliency}A)\cite{ZhaopingBook2014}.  
V1 is unique, however, in that lesioning it alone abolishes visually 
guided saccades for weeks\cite{IsaYoshida2021}. 
The fastest set of output axons from visual cortex are those
projecting to SC  \cite{FinlayEtAl1976,NowakBullier1997}, and
V1 contributes most of these axons \cite{CerkevichEtAl2014}.

According to V1SH \cite{LiTICS2002},  the saliency map is represented by the highest 
V1 response to each visual location (without decoding scene properties).
Intracortical V1 interactions cause nearby neurons tuned to similar features to 
suppress each other \cite{ZhaopingBook2014}.  For example, 
when an image contains a uniquely oriented bar among uniformly oriented background bars,
responses to the background bars suffer from iso-orientation suppression \cite{KnierimVanEssen1992},
whereas the response to the orientation singleton escapes this suppression.
When monkeys searched for such a  singleton (Fig. \ref{fig:FigV1Saliency}B-1),
trial-by-trial fluctuations in the initial neural responses to the singleton 
(at a latency $l=$ 40--60 ms after visual input onset) were inversely correlated with 
saccade latencies to the singleton \cite{YanZhaopingLi2018}.
Hence, these initial responses are saliency signals (read out by SC, Figure \ref{fig:FigRose}).
Also, electrical or optogenetic activations of monkey V1 evoke saccades towards 
the retinotopically corresponding locations\cite{TehovnikEtAl03, JazayeriEtAl2012}.
In monkey V4, saliency signals likely from V1 emerge at $l\approx $ 60 ms in layers not 
receiving feedback\cite{WesterbergEtAl2023}.
In FEF and parietal cortex, responses do not distinguish 
salient from non-salient items until $l\ge $ 100 ms \cite{ThompsonEtAl1996, ConstantinidisSteinmetz2001}.

Direct retinal inputs to SC appear insufficient for SC visual responses 
or for guiding saccades in normal monkeys.
Muscimol inactivation of lateral geniculate nucleus (LGN), which relays retinal inputs 
to V1, abolishes any visually evoked SC responses\cite{YuEtAl2024} (Fig. \ref{fig:FigV1Saliency}C-2).  
Equally, in a study \cite{IsaYoshida2021}, V1 was lesioned unilaterally after monkeys had learned to earn reward by
making a saccade to a light spot presented randomly at one of the four corners of a display (Fig. \ref{fig:FigV1Saliency}C-1).  
Afterwards, when the light spot was at either corner 
of the affected hemifield, saccades were aimed randomly into this hemifield, 
apparently  memory guided (mediated by FEF) to bring the target 
into the intact visual field, or to earn reward by chance.  
Blindsight after V1 lesion emerges only after months or years of neural reorganization \cite{IsaYoshida2021}.
Note that another study \cite{WhiteEtAl2017} reported that 
saliency signals in monkeys emerged in SC ($l\approx 65$ ms) before they 
did in V1 ($l\approx 139$ ms).  Replication studies should reconcile discrepancies.
Meanwhile, SC in lower vertebrates does compute saliency\cite{Zhaoping2016Evolution, WuXuEtAlLi2025}.

A  hallmark of V1's role is the gaze capture by a singleton  unique in  the eye of origin of visual 
inputs \cite{Zhaoping2008OcularSingleton, ZhaopingPeripheral2024}.
For example (Fig. \ref{fig:FigV1Saliency}D-1), the reaction time (RT) to find a unique 
letter {\sf 'T'} among letter {\sf 'L'}s normally increases with the number $N$ of {\sf 'L'}s; however, 
if the {\sf 'T'} is shown to one eye and the  {\sf 'L'}s are shown to 
the other eye, the {\sf 'T'}  pops out by 
a short RT independent of $N$ \cite{Zhaoping2008OcularSingleton}.
V1 is unique among cortical areas in having a substantial percentage of monocular neurons (in the binocular visual field), encoding the eye-of-origin.
Normal observers cannot discriminate eye-of-origin,
suggesting that the eye-of-origin information is lost beyond V1 (Fig. \ref{fig:FigCPD}A). 
Moreover, pop-out by eye-of-origin can dominate pop-out by orientation
in a search for an orientation singleton bar.  
When a non-target bar is an eye-of-origin singleton (Fig. \ref{fig:FigV1Saliency}D-1), 
it tends to capture the first gaze shift during search when both singletons are sufficiently eccentric \cite{Zhaoping2008OcularSingleton, ZhaopingPeripheral2024}.
This gaze distraction exemplifies looking without seeing (the eye-of-origin) \cite{ZhaopingPeripheral2024}.

Without free parameters, V1SH quantitatively predicted  a distribution of the RTs 
to find a feature singleton unique in color (C), orientation (O), and
motion direction (M) from distributions of the RTs for finding a singleton 
unique in only one or two of these feature dimension (Fig. \ref{fig:FigV1Saliency}D-2).
The prediction was confirmed \cite{ZhaopingZhe2015}. 

Although endogenous control can readily suppress salient distractors when their
features or locations are predictable \cite{IpataEtAl2006,GaspelinEtAl2025}), 
suppression is delayed for unpredictable distractors.
For example (Fig. \ref{fig:FigV1Saliency}B-2), 
when monkeys searched for a shape-defined 
target while a color singleton distractor was unpredictable in
color, shape, and location \cite{KlinkEtAl2023},
V4 neural responses to the distractor were initially higher 
than responses to the other items before they were 
suppressed at $l\gtrsim 100$ ms.  These responses likely 
reflect early feedforward inputs from V1 followed by
top-down suppression from, e.g., FEF and parietal 
cortex\cite{ThompsonEtAl1996, BuschmanMiller2007, BisleyGoldberg2010, ZhouDesimone2011}.
Long-latency target enhancement and distractor suppression are also evident 
in magnetoencephalography imaging of human V1 responses\cite{DueckerEtAl2025}.

Patients with parietal lesions have impaired attention \cite{BisleyGoldberg2010}.
However, during their search for a vertical bar among uniformly tilted oblique bars 
and a distracting horizontal bar (Fig. \ref{fig:FigV1Saliency}D-3), 
the distractor tends to capture their short-latency first saccade 
even when both the target and distractor are in the 
affected hemifield\cite{DombroweEtAl2012}.  
Therefore, parietal cortex does not control bottom-up saliency, 
whose effects are typically short-lived \cite{DonkVanZoest2008}.
Indeed, data from high-field functional magnetic resonance imaging (fMRI) 
suggest that saliency signals in parietal cortex originate 
from V1 \cite{LiuEtAlZhaopingZhang2025}. 

Subsequently, object meaning often predicts gaze shifts better than saliency \cite{StollEtAl2015}, consistent with 
the short-lived nature of saliency effects \cite{DonkVanZoest2008}.
After initial saliency-driven saccades extract scene structure, 
subsequent gaze shifts can be endogenously influenced by meanings. 
Thus, patients with visual object agnosia show normal early, but altered later, 
gaze patterns \cite{MannanEtAl2009}.
However, initial fixations set initial conditions for subsequent 
explorations, and gaze position can shape perceived meaning, 
as in viewing an ambiguous Necker cube in Fig. \ref{fig:FigV1Saliency}A-2 \cite{MengTong2004,HsuChen2025}. 
Whether brief saliency exerts a lasting ``butterfly effect" on subsequent 
explorations remains unclear.

\section*{Visual recognition through a bottleneck via feedforward from V1 
and feedback to V1}

\begin{figure}[p!]
\centering
\includegraphics[width=\textwidth]{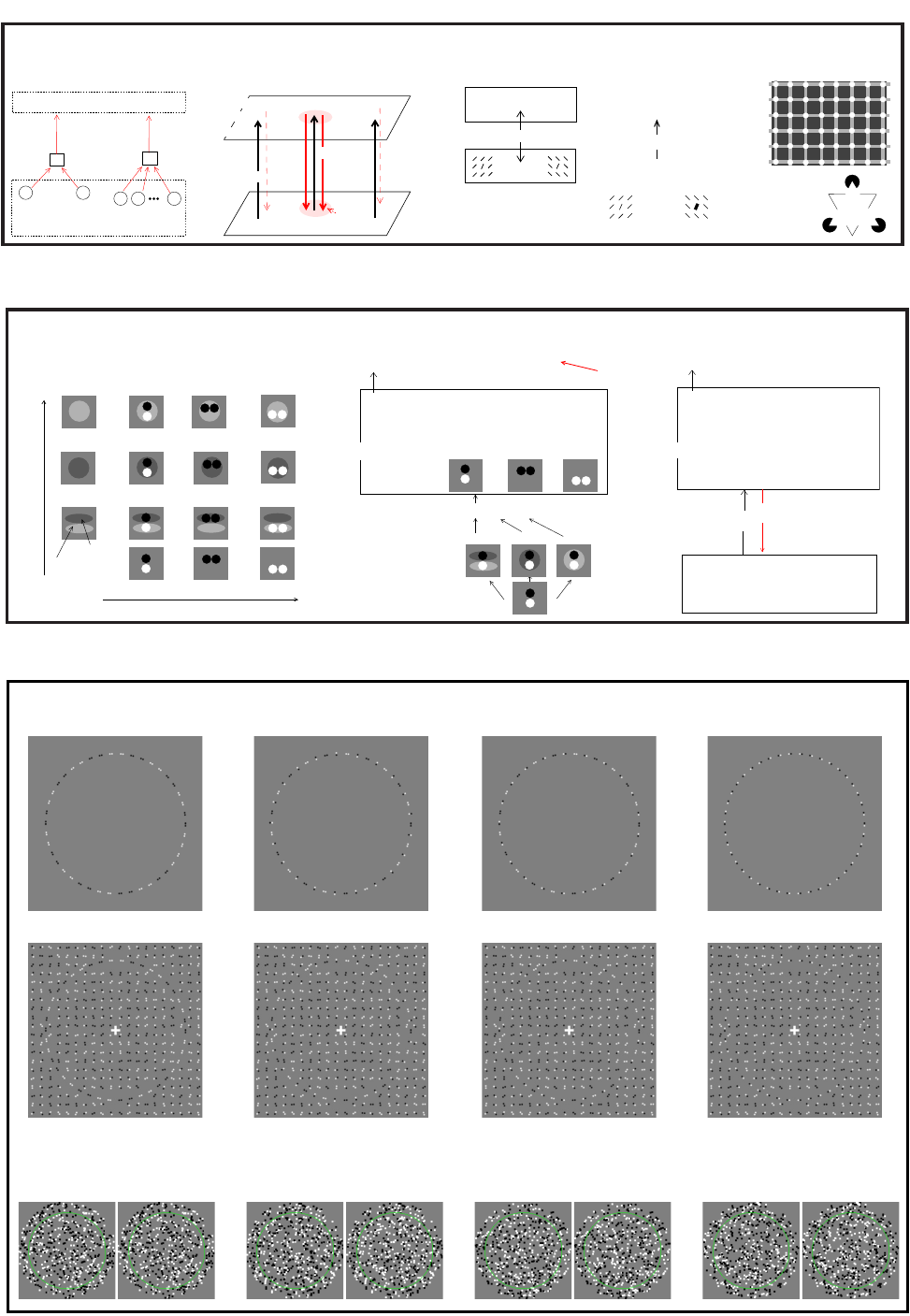}
\caption{\label{fig:FigCPD}, caption in the next page}
\end{figure}
\begin{figure}[p!]
\ContinuedFloat
\caption{Caption of figure \ref{fig:FigCPD} in the preceding page:
\small V1 works with higher brain areas to realize recognition under a processing bottleneck.
A: overview of the CPD theory and its account of some perceptual phenomena.
A-1: the bottleneck starts at V1's output to downstream areas, discarding
information such as eye-of-origin. Feedback, directed mainly to the
central field representation, queries for additional information to support 
ongoing recognition.
A-2: visual crowding arises from a lack of the feedback query
in the peripheral visual field to aid recognition using impoverished information 
through the bottleneck.  Salient items evoke stronger V1 responses and are 
less susceptible to information loss \cite{KimPasupathy2024}.
A-3: the lack of the query also makes the peripheral field vulnerable
to illusions as errors in recognition, e.g., the illusory scintillation 
is visible only outside the central gaze.  
The white Kanizsa triangle is more vivid in the central than the peripheral    
field: the feedback query aids perceptual completion by analysis-by-synthesis. 
B: the flip tilt illusion is predicted in the peripheral field, if the bottleneck 
blocks the V1 signals, e.g., neural responses $r_b$ and $r_c$,  necessary
to disambiguate a hetero-pair of dots from homo-pairs of dots orthogonal to the 
hetero-pair. This illusion makes a hetero-pair of dots appear tilted 
orthogonal to its actual tilt (of
the axis of alignment between the two dots). This illusion can be vetoed by feedback to 
V1 querying for the relevant information.
C: demonstration of CPD predicted flip tilt illusion and the reversed depth illusion.
C-1: The hetero-pairs on each ring are in the peripheral visual field when gaze is
at the center of the ring, and should appear --- by the illusion ---  tilted 
orthogonal to their true orientation.  This makes the ring's visibility in noise drop 
from C-1b to C-1c to C-1d.  All the homo-pairs on the rings are parallel to the tangent of the ring.  
C-2: C-2a to C-2d are stereograms (a left-eye image and a right-eye image each)
analogous to C-1a to C-1d, homo- and hetero-pairs are dichoptic pairs of dots (one dot each eye) 
to signal depth, veridically and illusorily (as reversed-depth), by V1 neurons tuned to binocular 
disparity. 
Each stereogram contains a disk (made of homo- and/or hetero-pairs, all the homo-pairs have disparity $d$) 
and a surrounding ring (made of zero-disparity homo-pairs).   
Free-fuse each stereogram in the central visual field to view.
C-2a reveals a disk (disparity $d$) veridically in front of the ring.  
(The zero-disparity green circles mark the inner borders of the rings for illustration only.)
In C-2d, the disk's depth is indiscernible, since the feedback query vetoes the 
illusory signals from the hetero-pairs.  The disk appears in front of the ring more 
clearly in C-2b than in C-2c, demonstrating that in C-2b, 
our brain constructively uses V1's illusory depth signals (from the hetero-pairs) that 
agree with the veridical and verifiable depth signals from the homo-pairs \cite{ZhaopingHeteroHomoDots_RDS2021}.
}
\end{figure}

\FloatBarrier

V1's role in looking motivates the idea \cite{ZhaopingNewFramework2019} that 
the information bottleneck starts immediately at V1's output 
to downstream stages \cite{ZhaopingNewFramework2019} (Figure \ref{fig:FigCPD}A). 
For example, a binocular downstream neuron preferring vertical orientation 
could pool inputs from two monocular V1 neurons of different eyes, 
discarding eye-of-origin information. 
More generally, downstream combination of V1 signals can 
lose substantial details; when mainly  summary statistics of visual inputs are preserved,
phenomena like  visual crowding emerge 
\cite{WhitneyLevi2011, RosenholtzEtAl2019, ZhaopingPeripheral2024}.

By the Central-Peripheral Dichotomy (CPD) theory \cite{ZhaopingNewFramework2019}, 
peripheral and central visual fields are mainly for looking and seeing \cite{ZhaopingBook2014, Nuthmann2014}, 
respectively.
For seeing in light of the bottleneck, the CPD theory proposes that downstream stages 
query for additional relevant information for the ongoing processing by
sending feedback to upstream areas like V1. This feedback focuses on the central field 
representation.

Lacking this feedback, the peripheral field suffers from visual crowding 
\cite{ZhaopingNewFramework2019, ZhaopingPeripheral2024}, 
see Figure \ref{fig:FigCPD}A-2.  
Thus a peripheral bar that is surrounded by 
other bars is  more legible when it has 
a larger orientation contrast from the other bars and so 
is  salient.
Such a salient bar evokes a higher V1 response by escaping iso-orientation 
suppression, making its information better preserved through the bottleneck.
Indeed, salient inputs are more faithfully represented in V4 responses \cite{KimPasupathy2024}.

Besides crowding,  information impoverishment can lead to visual illusions. 
For example, in Figure \ref{fig:FigCPD}A-3,
illusory scintillations are only visible in white disks outside the central gaze; 
they are vetoed in the central visual field by the feedback query. 

CPD predicts the flip tilt illusion in the peripheral visual field (Figure \ref{fig:FigCPD}B).
Three different pairs of dots ---
a vertical hetero-pair of dots (one black, one white), a horizontal homo-pair of black dots, or a horizontal homo-pair of white dots ---
evoke three different patterns of responses  $(r_a, r_b, r_c)$ from V1 neurons $a$, $b$, and $c$.
Neuron $a$ prefers horizontal orientations, but is excited
by all three pairs, including the vertical hetero-pair since this pair's black and white 
dots fall into this neuron's off- and on-subfields.

During recognition, downstream areas entertain  hypotheses $H_1$, $H_2$, and $H_3$ 
that the retinal input is the hetero-pair, the black homo-pair, or the
white homo-pair, based on V1 responses admitted into the bottleneck.
If the vertical hetero-pair is shown but only $r_a$ is admitted, 
perception is ambiguous: all three hypotheses can produce $r_a>0$ (Figure \ref{fig:FigCPD}B-2).  
Perceived orientation is likewise ambiguous, with $H_2$ and $H_3$ voting 
horizontal and $H_1$ vertical, so a forced-choice
report is likely horizontal from the majority voting.
Thus, when the bottleneck admits only $r_a$, a retinal hetero-pair 
appears tilted orthogonally to the actual tilt --- the flip tilt illusion.
Formally, substantial likelihood $p(r_a|H_i)$ of $r_a$ 
given $H_i$ and prior probability $p(H_i)>0$ for $H_i$ yield 
comparable posterior $p(H_i|r_a)$ between $i$'s, 
and orientation inference follows
$p(\textrm{orientation} | r_a) =\sum_i p(\textrm{orientation} |H_i) p(H_i|r_a)$.

If downstream areas send feedback to V1 to query for $r_b$ and/or $r_c$,
responses $r_b=0$ and $r_c=0$ to the hetero-pair 
veto hypotheses $H_2$ and $H_3$, leaving only $H_1$.
This removes the ambiguity and vetoes the illusion.  
Such a query requires brain's internal knowledge  ($p(r_b, r_c|H_i)$) 
of expected $(r_b, r_c)$  for each $H_i$, and thus the knowledge that ($r_b, r_c)$ can discriminate between the hypotheses.

The Feedforward-Feedback-Verify-reWeight (FFVW) algorithm has been proposed
to perform recognition \cite{ZhaopingNewFramework2019} (Figure \ref{fig:FigRose}). 
The bottleneck first admits feedforward responses $\br$,
generating hypotheses $H_i$'s with weights $w_i$ (which should 
grow with $p(H_i|\br )$).  Then, feedback queries for additional  responses $\br'$.
In our example, $\br = r_a$ proposes the hypotheses $H_i$'s, 
and $\br' = (r_b, r_c)$ disambiguate them.  
Using learned expectations
of would-be responses $\hat \br' (H_i)$ for each $H_i$, 
the brain compares the synthesized $\hat \br' (H_i)$ with the actual 
$\br'$ to reweight each hypothesis: $w_i$ increases if  $\hat \br' (H_i)\approx \br'$ 
and decreases otherwise.  
Here $(r_b, r_c) = (0, 0)$ boosts $w_1$ and suppresses $w_2$ and $w_3$.
This FFVW algorithm thus implements analysis-by-synthesis.

Figure \ref{fig:FigCPD}C-1 demonstrates the flip tilt illusion. 
In noise, the C-1d ring is the least conspicuous in our peripheral field.
This is because it only contains hetero-pairs which appear illusorily orthogonal to the tangent.
This illusion also makes the C-1b ring more conspicuous than the C-1c ring in the 
peripheral field.  However, when gaze traces the circumferences of the rings,
the feedback query in the central visual field vetoes the illusion,
making the C-1b ring appear less smooth than the C-1c ring.

The effects of central-field feedback are also apparent in
random dot stereograms in Figure \ref{fig:FigCPD}C-2.
Dichoptic homo-pairs arise from physically plausible three-dimensional (3D) dots.
However, a dichoptic hetero-pair lacks a  realistic 3D interpretation 
(since a black dot in one eye is paired with a white dot in the other). 
It excites a V1 neuron preferring a binocular disparity
opposite in sign to the true interocular disparity \cite{Zhaoping_iScience2025},
analogous to the excitation of horizontal-preferring neurons by vertical 
hetero-pairs in Figure \ref{fig:FigCPD}B.
This results in the reversed depth illusion \cite{ZhaopingAckermann2018}, 
an analog of the flip tilt illusion.

In Figure \ref{fig:FigCPD}C-2, C-2a is a standard random dot stereogram 
depicting a disk of binocular disparity $d$ in front of 
a zero-disparity surrounding ring.  In C-2d, all the disk dots are hetero-pairs 
of disparity $d$. Their V1 responses signal a disk behind the ring. 
However, free fusion of C-2d in the central visual field 
yields no clear depth ordering, because the feedback query 
(for, e.g., the monocular signals in V1) vetoes the misleading V1 signals.  
(V4 neurons are indeed less sensitive to the reversed depth 
signals \cite{TanabeEtAl2004}.)
In the peripheral visual field, which lacks the feedback query, 
the C-2d disk appears illusorily behind the ring as predicted \cite{ZhaopingAckermann2018}. 
As predicted \cite{Zhaoping_iScience2025}, 
in the central visual field, interrupting the feedback by backward masking renders 
the illusion visible (and similarly makes the flip tilt illusion visible). 


In Figure \ref{fig:FigCPD}C-2bc, the hetero-pairs make up 50\% of the disk dots.
Their disparity is opposite to (in C-2b) or the same as (in C-2c) the disparity
of the homo-pairs. Hence, their evoked V1 responses reinforce
or oppose the veridical depth signals from the homo-pairs.
Consequently, in the central visual field, 
the disk appears more clearly in front in C-2b than in C-2c. 
Hence,  our brain constructively exploits illusory or ambiguous feedforward signals which align with verifiable 
veridical cues \cite{ZhaopingHeteroHomoDots_RDS2021}.  
This likely explains why the white Kanizsa triangle\cite{Kanizsa1976} in Figure \ref{fig:FigCPD}A-3 
appears more vivid in the central than the peripheral visual field \cite{MooreEtAl2025}:
this triangle is (part of) a perceptual hypothesis suggested by fragmentary retinal 
inputs, the feedback query confirms it by verified occlusions of the black disks and another triangle.

Retrograde tracing in marmoset V1 shows that feedback axons are denser, by an order of magnitude, 
in V1's representation of the central, than the peripheral, visual field \cite{MajkaZhaopingRosa_VSS2026}, confirming the critical CPD prediction.  
Consistent with this, frontal areas exhibit stronger functional connectivity with central 
than peripheral V1 \cite{SimsEtAl2021}, and feedback from V4 to V1 inferred from  
Granger causality analysis of neural activities is concentrated in foveal V1\cite{MoralesEtAl2024}.
Equally, CPD predicts that indications of top-down feedback under ambiguous or 
difficult perceptual conditions \cite{KarDiCarlo2021, XinEtAl2025, PizzutiEtAl2025} 
should decrease with increasing visual field eccentricity.

By CPD, a visual percept or phenomenon that is 
stronger in the central versus peripheral visual field is associated with a rich versus absent feedback query \cite{ZhaopingNewFramework2019}.
Accordingly, surface percepts arising from 
surface completion, 3D shape from 2D geometry, surface transparency, or 
closure --- which are stronger in the central than the peripheral 
visual field \cite{MooreEtAl2025} --- are predicted to rely
on feedback queries.  
Crowding is predicted to appear in the central visual field if the feedback is disrupted 
(e.g., by backward masking like in \cite{Zhaoping_iScience2025}). 
Likewise, foveal feedback is predicted to be weaker for amblyopic individuals (since they exhibit foveal crowding) \cite{Levi2020, Hess2025}.

Whereas  feedback queries adapt to ongoing recognition, 
I conjecture that the feedforward connections driving the initial 
sweep, particular from V1, are less flexible, especially in the peripheral 
visual field.  These connections apparently favor letting salient inputs 
through the bottleneck.  They are likely shaped during development and 
evolution to optimize initial recognition across scenes, 
and are likely modifiable to a limited extent in adulthood by perceptual 
learning\cite{LuDosher2022} or by task and context\cite{LiEtAl2004}.

\section*{Discussion and outlook}

V1's functions are tied to the looking-and-seeing framework  \cite{ZhaopingBook2014}, in which percepts 
are integrated across fixations via brisk 
saccades \cite{MelcherColby2008, LiangZhaoping2024, KroellRolfs2025} 
while the visual world remains relatively stable \cite{GolombMazer2021}.  
Because saccades bring peripheral objects to the fovea, feedback queries can
link visual signals $\br$ at a peripheral location to 
the foveally queried signals $\br '$ at the next moment 
through ongoing perceptual hypotheses $H_i$'s about the scene.
This perspective aligns with evidence of feedback to fovea V1 
from peripheral stimuli prior to saccades and of impaired peripheral 
recognition  by foveal masking \cite{WilliamsEtAl2008, KnapenEtAl2016, FanEtAl2016, OlettoEtAl2026}. 
Likewise, fixational eye movements can be understood as part of the feedback query process 
rather than random gaze fluctuations \cite{RucciPoletti2015}.

The perceptual hypotheses $H_i = \{ \theta_1, \theta_2, ... \}$ 
are parameterized by $\theta_1, \theta_2 ...$ 
that specify scene properties, e.g., the shape and color of objects in $H_i$.  
One parameter $\theta_j$ specifies object viewpoints.
In trans-saccadic recognition,
$\theta_j$ should be updated to generate the would-be 
visual signals $\hat \br'$ after the saccade (including for relevant or salient objects that are not the saccadic target),
using an efference copy of the saccadic command \cite{Wurtz2018}.  
This aligns the expected $\hat \br'$ with the actual $\br'$ after the saccade, 
supporting perceptual stability.  
This perspective is related to works on neural receptive field remapping 
across saccades (e.g., \cite{WangEtAlQian2024, CavanaghMelcher2026}). 
The bottleneck imposes a massive information reduction 
from retinal inputs to percepts ($H_i$'s), allowing 
perceptual information accumulated across fixations to
be maintained in capacity-limited working memory \cite{LuckVogel2013} before 
it is lost or consolidated into longer-term memory.
Precise eye-tracking and free viewing behavior \cite{YatesEtAl2023} 
are often needed to better understand natural vision, in which 
visual inputs are largely selected overly rather than covertly. 

Some falsifiable theoretical predictions (Table \ref{table:table1} ) remain to be tested.  
Our framework raises further questions, including
how the FFVW algorithm is implemented neurally; 
whether and how this algorithm and brain's internal states influence 
downstream activities (e.g., in V1);
how exogenous and endogenous control of gaze and attention 
are prioritized and integrated in the brain; 
how some of the information loss through the bottleneck helps achieve object invariance in recognition;
and what additional roles V1 plays in tasks such as navigation, balance, avoidance, and escape
--- behaviors that extend beyond looking-and-seeing since they require little or no 
detailed recognition \cite{SerenoHuang2014, VaterETAl2022, ZhaopingPeripheral2024}.
Addressing these questions connects our theories with other lines of 
research and can generate further falsifiable predictions.  
Furthermore, understanding primate vision should inform multisensory processing 
across species --- from primate V1 to the superior colliculus (optic tectum) 
in lower vertebrates \cite{Zhaoping2016Evolution} and from looking-and-seeing 
to multisensory orienting-and-recognition \cite{ZhaopingCPD2023} ---
as all animals face processing bottlenecks.

\begin{table}[h]
\caption{Some falsifiable predictions of the theories concerning primate V1's functional roles}
\small{
\centering
\begin{tabular}{|p{2.5 cm}|p{9cm}|p{2cm}|}
\hline
{\bf theory of V1's functions }& {\bf theoretical prediction} & {\bf experimental status} \\ \hline
\multirow{3}{*}
{\parbox{2.5cm}{%
\vspace {2mm}  V1 Saliency \\ Hypothesis (V1SH)}}
  & ocular singleton pops out to attract gaze and attention
  & confirmed\cite{Zhaoping2008OcularSingleton}, Fig. \ref{fig:FigV1Saliency}D-1  \\ \cline{2-3}
  & Higher V1 responses to a search target tend to precede a shorter latency saccade to the target
  & confirmed \cite{YanZhaopingLi2018}  Fig. \ref{fig:FigV1Saliency}B-1 \\ \cline{2-3}
  & zero-parameter quantitative distribution of RTs for 
	the feature singleton search task described in\cite{ZhaopingZhe2015}
  & confirmed \cite{ZhaopingZhe2015} Fig. \ref{fig:FigV1Saliency}D-2 \\  \cline{2-3}
  & a collection of other predictions 
  & confirmed and reviewed in \cite{ZhaopingBook2014}.  \\  \hline
\multirow{4}{*}
{\parbox{2.5cm}{%
\vspace {2mm}  Central-\\Peripheral \\
Dichotomy (CPD) \\
theory }}
  & the flip tilt illusion and the reversed depth illusion in the peripheral visual field
  & confirmed\cite{ZhaopingAckermann2018, ZhaopingPeripheral2024}, Fig. \ref{fig:FigCPD}C  \\ \cline{2-3}
  & the reversed depth illusion becomes visible in the central visual field by backward masking
  & confirmed \cite{Zhaoping_iScience2025}   \\ \cline{2-3}
  & feedback in the ventral visual pathway (for recognition) to V1 is
	mainly directed to the representation of the central visual field
  & confirmed \cite{MajkaZhaopingRosa_VSS2026, MoralesEtAl2024} \\  \cline{2-3}
  & cortical neurons (e.g., V2, V4, perhaps also V1) are more sensitive to the reversed feature signals by hetero-pairs (in, e.g., orientation,  depth)
	at shorter latencies after stimulus onset before the feedback query 
becomes effective, or at a more peripheral visual location 
 where the feedback query is weaker & to be tested \\ \cline{2-3}
  & in upstream areas such as V1 and V2, the percentage of simple cells is higher nearer to the foveal representation, since simple cell responses have
more sensory details for the feedback query
		 & to be tested \\ \cline{2-3}
  & feedback to V1 along the ventral pathway is more likely directed to the binocular rather than monocular cells \cite{ZhaopingNewFramework2019}
  & to be tested \\ \cline{2-3}
  & crowding appears in the central visual field when the feedback query is impaired, e.g., by backward masking 
  & to be tested \\ \cline{2-3}
  & amblyopic eyes, which exhibit crowding in the central visual field, are associated with weaker or absent feedback query 
  & to be tested \\ \cline{2-3}
  &  perceptual phenomena, such as perceptual organization for 
surface completion \cite{MooreEtAl2025}, that are 
	stronger in the central than peripheral visual field regardless of 
input spatial resolution should involve the feedback query, and can be weakened by impairing the feedback 
(e.g., by backward masking).
  & to be tested \\ \cline{2-3}
  & perceptual phenomena, such as crowding and some illusions, 
	that are stronger in the peripheral than the central  visual field 
regardless of input spatial resolution are associated with a weaker or a lack 
of the top-down feedback query.  
	& to be tested \\ \hline
\end{tabular}
\label{table:table1}
}
\end{table}
\FloatBarrier

\section*{Acknowledgement}  This work is supported in part by University of T\"ubingen and the Max Planck Society. I thank 
Peter Dayan for helpful comments to the manuscript. 



\end{document}